# A Modest Proposal:
# (Possible) Implications of Appropriation

Peter Kinnaird

**ABSTRACT**

As technologies are developed and constructed, designers may or may not be aware that they are embedding politics and values into their artifacts. Computer scientists operate and advance their field by building layers of abstraction into software and hardware to reduce the complexity of interfaces, making the artifacts they create *zuhanden* for others in part by imposing constraints. The increasing reliance of global populations and economies on communication mediated by many information and communications technologies (ICTs) transforms them from applications into communications infrastructure, elevating the importance of considering the values embodied in those infrastructures. I argue that the status quo bias and economic inertia of built-infrastructure requires a reevaluation of research in light of the de facto global technocracy to consider a nouveau social contract between infrastructural theorists, scientists, designers, and engineers and the current and future generations who will be constrained by that infrastructure as it is reified. The CHI community is uniquely situated to establish a norm of including a discussion of the implications of appropriation as first class topic in research output.

**INTRODUCTION**

The hope and promise of a great deal of scientific research, including theoretical computer science and HCI research, is that the ideas, discoveries, and designs expressed might one day be instantiated and used to alter the world [5]. Yet the discoveries expressed in scholarly research might be appropriated in ways unforeseen by the authors or contribute to systematic, unpredictable effects on society or the environment [8,9]. The responsibility for these effects does not ultimately rest on the shoulders of the scholar given the impossibility of conceptualizing all of the potential (mis)appropriations of their work, but on the implementation of technology informed by their research. Nevertheless, a small measure of responsibility remains with the scholar to consider the implications of appropriation of their discoveries.

Applications of Computer Science are built on the notion of layers of abstraction. Graphical User Interface (GUI) programmers are not generally concerned about making sure the mouse cursor moves when the user physically moves the mouse since the Operating System (OS) is handling that. Likewise, OS programmers are not generally concerned with translating network signals between electrical impulses and bits since that takes place in the Network Interface Card (NIC). If GUI programmers had to fully understand and implement every element of their GUI it would be virtually impossible to write even the simplest web browser, for example. Computer Scientists working on this underlying hardware and software infrastructure make it *zuhanden* for other programmers by authoring Application Programming Interfaces (APIs) and agreeing on standards (like x86) so that those other programmers can built on top of their contributions without fully needing to know the details of the implementation. Although powerful tools, APIs and layers of abstraction introduce countless opportunities for the introduction of values into computing infrastructure (regardless of the intent of the author) by constraining the space of possible uses by definition.

Every day global society becomes more and more dependent on computer mediated communication for everything from discovering romantic partners to automated financial oversight. The construction and maintenance of these communications technologies is typically overseen by a blend of industry, scientists, interested stakeholders (e.g. RFC's maintained by the IETF), and traditional governing bodies. The extraordinary influence wielded by the unelected, tech-savvy elite over this infrastructure suggests the de facto presence of a market- and meritocratically-elected global technocracy determining the values of global communications infrastructure. The emergence of this amorphous and dynamic global governing body requires the consideration of a nouveau social contract between contributors to the design and construction of infrastructure and the current and future generations who will be constrained by that infrastructure.

Given the challenges associated with making predictions, it is not the responsibility of each researcher to attempt to embed positional values in their work, but merely to consider the social and environmental arenas that could be significantly impacted by their work. The CHI community should establish a norm of including at least a very brief discussion of these *implications of appropriation* as a first class inclusion in research output.

**BACKGROUND**

**Values Free Science**
Max Weber famously advocated values-free science since value laden parables dressed up in the trappings of scholarly research might seduce the naïve. Although the consensus among scientists is certainly to deal in facts rather than values, two important questions remain: 1) is it possible for us to do so, and 2) do moral imperatives demand that we ought to conduct values-free science or agenda-driven research and presentation?

**Artifacts and Politics**

Langdon Winner resoundingly stated that all artifacts have politics, whether the strong politics of expressed intent from a grand architect or the softer politics of their interpretation [28]. Artifacts could be broadly defined to include the informational outputs of scholarly research, or more narrowly defined as built objects in the world. The impact of the values embedded in artifacts is exceptionally salient in the case of infrastructural artifacts. Winner's most memorable example is Robert Moses, a city planner in New York City, who decided to construct low bridges over Long Island expressways. Winner contends that this decision was made with the intent of maintaining segregation at Long Island parks by preventing public busses from using the expressway thus excluding those too poor to take their own car. Although the details of this story have been called into question by detractors, including Bernward Joerges, the story retains truth in the effects of the decision if not in the purpose behind it. Even Joerges, Winner's most well-known detractor recognizes that "the power represented in built and other technical devices is not to be found in *the formal attributes of these things themselves*. Only their *authorization*, their legitimate representation, gives shape to the definitive effects they may have [16]." Yet the very fact that the outputs of scholarly research form artifacts disallows the absolution of even theoretical research.

**Infrastructure's Uncertain Endurance**

Star suggests that realized infrastructure becomes ecological, inseparable from the built environment [26]. Once infrastructure is in place and technologies are built to suit the standards of that infrastructure, the infrastructure takes on substantial economic inertia and becomes increasingly difficult to modify, replace, or remove [1,19]. Given the economic inertia and human propensity for exhibiting status quo bias [18,24,27], we must consider the endurance of infrastructure uncertain.

**Future Stakeholders**

Researchers sometimes consider the impact of their research and its wide-scale deployment on differing global societies [2,22], but they rarely consider their own society's possible future values in this exercise. A notable exception is some recent theoretical design work from Latour. Latour considers the possibility that the politics of artifacts may shift since they exist in constantly shifting contexts [20].

Historian Tony Judt, writes concerning the use of taxes:

*…[M]ost taxation goes towards either paying off past debt or investing in future expenditures. Accordingly, there is an implicit relationship of trust and mutuality between past taxpayers and present beneficiaries, present taxpayers and future recipients – and of course future taxpayers who will cover the cost of our outlays today. We are thus condemned to trust not only people we don't know today, but people we could never have known and people we shall never know, with all of whom we have a complicated relationship of mutual interest* [17].

The same can clearly be said of research and infrastructural investment. As key players in the ongoing development of future communications infrastructure which may be deployed globally, we need to consider our position as de facto surrogate representatives of the current and future populace and take that responsibility seriously. Indeed, we might frame the responsibility as a nouveau social contract between the users and the theorists and builders comparable to political social contract theory. Given that we have no idea what future societies (let alone current ones) will truly value, we have a responsibility as researchers to acknowledge our biases and consider the societal and environmental implications of appropriating our work.

**APPROACHES**

The ACM's code of ethics which all members agree to uphold states a number of moral imperatives which apparently dictate that its members conduct values-laden work.

*As an ACM member I will…contribute to society and human well-being. The principle…affirms an obligation to protect fundamental human rights and to respect the diversity of all cultures. An essential aim of computing professionals is to minimize negative consequences of computing systems…When designing or implementing systems, computing professionals must attempt to ensure that the products of their efforts will be used in socially responsible ways, will meet social needs, and will avoid harmful effects to health and welfare. …[H]uman well-being includes a safe natural environment. Therefore, computing professionals who design and develop systems must be alert to, and make others aware of, any potential damage to the local or global environment* [7].

A number of imperatives are provided in this statement that expressly apply to the designers and implementers of systems. Before examining the imperatives individually, we must consider the question of their application to researchers who are not designing or developing systems. I contend that the imperatives apply equally to all parties since one of the goals of even theoretical research is to eventually see application.

Others have suggested that HCI researches should pursue the direct and value-laden goal of promoting peace [14,15,25]. Value Sensitive Design is another, similar approach which suggests that we should develop technology that "we can and want to live with [10,11,12]." Still others have suggested that the Universal Declaration of Human Rights (UDHR) is a globally acceptable document from which we can draw values [2,23]. Attempting to embed the values of a document like the UDHR in infrastructure is ultimately doomed due to the reinterpretability of the UDHR and the right of each nation to regulate the rights therein described. Unlike source code, legal systems and documents, including the UDHR, are nearly always subject to human interpretation and reinterpretation [21]. For example, Article 5 of the UDHR states that everyone has the right to freedom of peaceful assembly and association yet, in the United States, for example, this right is regulated. Citizens wishing to peacefully assemble on a public road must usually obtain a permit. In some cases this right is waived subject to the judgment of appointed officials.

Philosopher James Carse describes the concept of finite and infinite games [6,13]. Finite games are things like reproducing geometry proofs or Conway's Game of Life in which there are a certain set of moves. We could also agree on the meaning of winning and losing, though those definitions might vary from one version of the game to the next. Infinite games are things like the US Constitution which can always be amended and reinterpreted. There is no way to 'win' with the constitution, only rules that we agree to follow (or break) with relatively open-ended possibilities. If Carse's thesis is correct, societal values are not only inherently unpredictable and unscripted, but also the central feature that enables society to grow and develop. As such, any statement of human values, however universal it may appear, is subject to change.

**RESEARCH AS INFRASTRUCTURE**
As researchers, we are rarely able to present the full picture of our work in a paper or presentation. In fact, we examine the data and contributions of our work and digest these into a sensible narrative that we can easily communicate to others. Presenting an absolutely complete view of our work would be impractical in most research papers since readers would need to comprehend every line of source code, every statement from a participant interview, every preliminary design sketch, and every variable and data point collected to make sense of the output. Even if these data were supplied with research papers, most researchers would still prefer to consume a simpler narrative that discards irrelevant alternatives and details explored by the author. In this sense, we can view the research paper as the interface to our work. We rely on the authors to supply sufficient caveats, scoping conditions, and descriptions of generalizability that we require to reuse or build on the work. These components are judgments of the researcher. To dampen the impact of values leaking into the research paper, we pursue rigorous academic programs and peer review coupled with editorial control. Yet no matter how much we dislike the idea that values are embedded in research papers, we require that they are in order to advance a field of science. Without those judgments every scientist would need to replicate science in order to understand that work and move forward.

Therefore, the research we produce is a kind of infrastructure on which further research is built. Although we occasionally shift paradigms and discount earlier research, we can expect that months or years from now, someone might consume our paper and take it for granted that we accurately and adequately conveyed what they need to move forward, just like the NIC converts pulses of electricity into bits so that the OS can work only in bits without concern for the modulation and demodulation required for conversion of bits to electricity. Just like any good API author, it is our responsibility, therefore, as authors to not only provide sufficient caveats to our work, but to consider the implications of appropriation.

**IMPLICATIONS FOR APPROPRIATION**
Democratization of knowledge resources means that 'anyone' might pick up our research and build on it. We have a minimal responsibility to consider areas that might be impacted if/when our research is used by others. Even if not a top-level section in a research paper, we should devote at least a single sentence to the consideration of social and environmental areas that might be impacted if our work was appropriated. For example, a research paper presenting the Map/Reduce algorithm might warn that mapping to parallel computational units implies an array of comparable machines standing by for computation. There are important environmental considerations to adopting the algorithm with respect to energy consumption. Future internet architecture descriptions ought to warn that adoption might impact user privacy at the expense of optimization and ubiquitous computing researchers might warn that large-scale deployment of smart phones could impact driving safety and in-person communications.

In the spirit of the thesis of this paper, I briefly consider the implications of appropriating the implications of appropriation as a first class inclusion in CHI research outputs. There is little doubt that such a norm would require researchers to maintain some degree of visibility into numerous fields. This requirement might slow the progression of highly specialized research. On the other hand, researchers have also shown that those who bridge structural holes are well-placed to advance their careers [3,4]. Such a section would also consume at least one line in most research papers. There is also the possibility that future implementers of a research paper may take it for granted that all possible implications of appropriation are listed in a paper when in reality a list may be incomplete.

The CHI community, with its characteristic blend of disciplines, is uniquely situated to consider these kinds of impacts. Although the authors of a paper contributing an internet architecture, for example, might not be qualified to fully characterize the nature of the impact of their design on privacy, they should at least maintain enough visibility into the necessary domains to recognize the possibility of an important impact.

**REFERENCES**
1. Brand, S. *How buildings learn: what happens after they're built*. Penguin Books, 1995.
2. Brown, I., Clark, D.D., and Trossen, D. Should specific